\documentclass[twoside,12pt,a4paper]{report}
\usepackage{amssymb}
\usepackage{natbib}
\usepackage{epsfig}
\usepackage{graphicx}
\usepackage{wallpaper}

\begin{document}
%\LLCornerWallPaper{1}{bg_footer.png}
\CenterWallPaper{0.8}{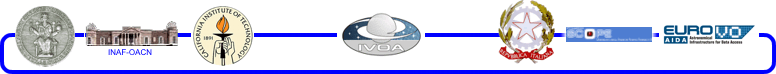}
\setlength{\wpYoffset}{-13cm}
\pagestyle{headings}

\begin{titlepage}

\begin{picture}(100,0)(-10,-10)
      \includegraphics[width=12cm]{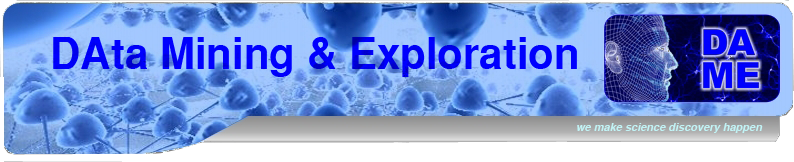}
    \end{picture}

  \begin{center}

    \vspace{10mm}
    \begin{figure}[htbp]
      \begin{center}
      \vspace{10mm}
      \epsfig{file=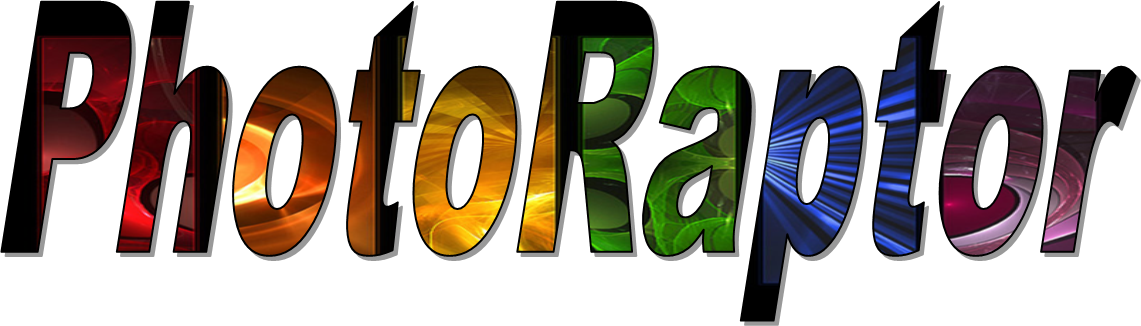,width=\textwidth}
      \end{center}
    \end{figure}
   \vspace{5mm}
  {\large \bf Photometric Research Application To Redshifts}\\

  \vspace{8mm}
  {\Huge \bf User Manual}\\
  \vspace{5mm}
  {\Huge \bf Release 1.2}\\
  \end{center}
  \vspace{28mm}
    {\large \bf S. Cavuoti$^1$, M. Brescia$^1$, V. De Stefano$^2$, G. Longo$^2$}\\

  \vspace{5mm} 
  \noindent {\large \bf \textit{1 - INAF - Astronomical Observatory of Capodimonte}}\\
  {\large \bf \textit{2 - University Federico II of Napoli}}\\

\pagestyle{empty}

\end{titlepage}

\newpage
\setcounter{tocdepth}{4}
\tableofcontents
\newpage

%%%%%%%%%%%%%%%%%%%%%%%%%%%%%%%%%%%%%%%%
% CHAPTER 0
\chapter{Introduction}
\label{photoraptor}
%%%%%%%%%%%%%%%%%%%%%%%%%%%%%%%%%%%%%%%%

Measurements for fainter objects are nowadays accessible by photometry thanks to the improving telescope technology. This makes photometric
redshift (photo-z) extremely attractive for observing programmes depending on redshift especially with the advent of modern panchromatic
digital surveys. In fact, future large-field public imaging projects, such as KiDS\footnote{\url{http://www.astro-wise.org/projects/KIDS/}}
(Kilo-Degree Survey), DES (Dark Energy Survey, \citealt{des2005}), LSST (Large Synoptic Survey Telescope, \citealt{ivezic2009}) and Euclid
\citep{euclidredbook} require extremely accurate photo-z to obtain accurate measurements that do not compromise the survey's scientific goals.\\

Due to the necessity to evaluate photo-z for a variety of huge sky survey data sets, it seemed important to provide the astronomical community
with an instrument able to fill this gap. Besides the problem of moving massive data sets over the network, another critical point is that a
great part of astronomical data is stored in private archives that are not fully accessible on line. So, in order to evaluate photo-z it is
needed a desktop application that can be downloaded and used by everyone locally, i.e. on his own personal computer or more in general within
the local intranet hosted by a data center.\\

The name chosen for the application is \textbf{PhotoRApToR}, i.e. \textbf{Photo}metric \textbf{R}esearch \textbf{Ap}plication
\textbf{To} \textbf{R}edshift (\citealt{cavuoti2015,cavuoti2014,brescia2014b}).\\
It embeds a machine learning algorithm and special tools dedicated to pre- and post-processing data. The ML model is the MLPQNA (Multi Layer
Perceptron trained by the Quasi Newton Algorithm), which has been revealed particularly powerful for the photo-z calculation on the base of a
spectroscopic sample (\citealt{cavuoti2012b,brescia2013a,brescia2013b,biviano2013}).\\

In order to favor the portability of this tool, the Graphical User Interface was developed in Java\footnote{\url{http://www.oracle.com/technetwork/java/index.html}} language and runs on top of a standard Java Virtual Machine (JVM), thus permitting its execution in a platform-independent way.

Main features of the presented application can be summarized as follows:

\begin{itemize}
\item \textit{Data table manipulation}: in order to navigate throughout user's data sets and related \textit{metadata}, as well as to prepare data tables to be submitted for experiments, there are several options to perform the editing, ordering, splitting and shuffling table rows and columns. A special set of options is dedicated to the missing data retrieval and handling, for instance Not-a-Number (NaN) or not calculated/observed parameters in some data samples;
\item \textit{Classification experiments}: the user can perform general multi-class classification problems, i.e. automatic separation of an ensemble of data by assigning a common label to an arbitrary number of their subsets, each of them grouped on the base of a hidden similarity. The classification here is intended as \textit{supervised}, in the sense that there must be given a subsample of data for which the right label has been previously assigned, based on the \textit{a priori} knowledge about the treated problem. The application will learn on this known sample to classify all new unknown instances of the problem;
\item \textit{Regression experiments}: the user can perform general regression problems, i.e. automatic learning to find out an embedded and unknown analytical law governing an ensemble of problem data instances (patterns), by correlating the information carried by each element (features or attributes) of the given patterns. Also the regression is here intended in a \textit{supervised} way, i.e. there must be given a subsample of patterns for which the right output is \textit{a priori} known. After training on such KB, the program will be able to apply the hidden law to any new pattern of the same problem in the proper way;
\item \textit{Photo-z estimation}: within the \textit{supervised} regression functionality, the application offers a specialized toolset, specific for photometric redshift estimation, able to learn the hidden correlation between photometric and spectroscopic information on a subset of sky objects (patterns of the KB), for which the spectroscopic redshift is available. After training, the system will be able to predict the right photo-z value for any new sky object belonging to the same type of KB;
\item \textit{Data visualization}: the application includes some $2D$ and $3D$ graphics tools, for instance multiple histograms, multiple $2D$/$3D$ scatter and line plots. Such tools are often required within astrophysical problems to visually analyze and explore data distributions and trends, as well as resulting from data mining experiments;
\item \textit{Data statistics}: all classification and regression experiments provide a statistical report about their output. In the first case, the typical confusion matrix \citep{stehman1997} is given, including related statistical indicators such as \textit{classification efficiency}, \textit{completeness}, \textit{purity} and \textit{contamination} for each of the classes defined by the specific problem (see Sect.~\ref{confmat} for details). For what the regression is concerned, the application offers a dedicated tool, able to provide several statistical relations between two arbitrary data vectors (usually two columns of a table), such as average, standard deviation ($\sigma$), Root Mean Square (RMS), Median Absolute Deviation (MAD) and the \textit{Normalized} MAD (NMAD, \citealt{hoaglin1983}), the latter specific for the photo-z quality estimation, together with percentages of \textit{outliers} at different multiples of $\sigma$ \citep{brescia2013b, ilbert2009}.
\end{itemize}

\begin{figure*}
\centering
\includegraphics[width=\textwidth]{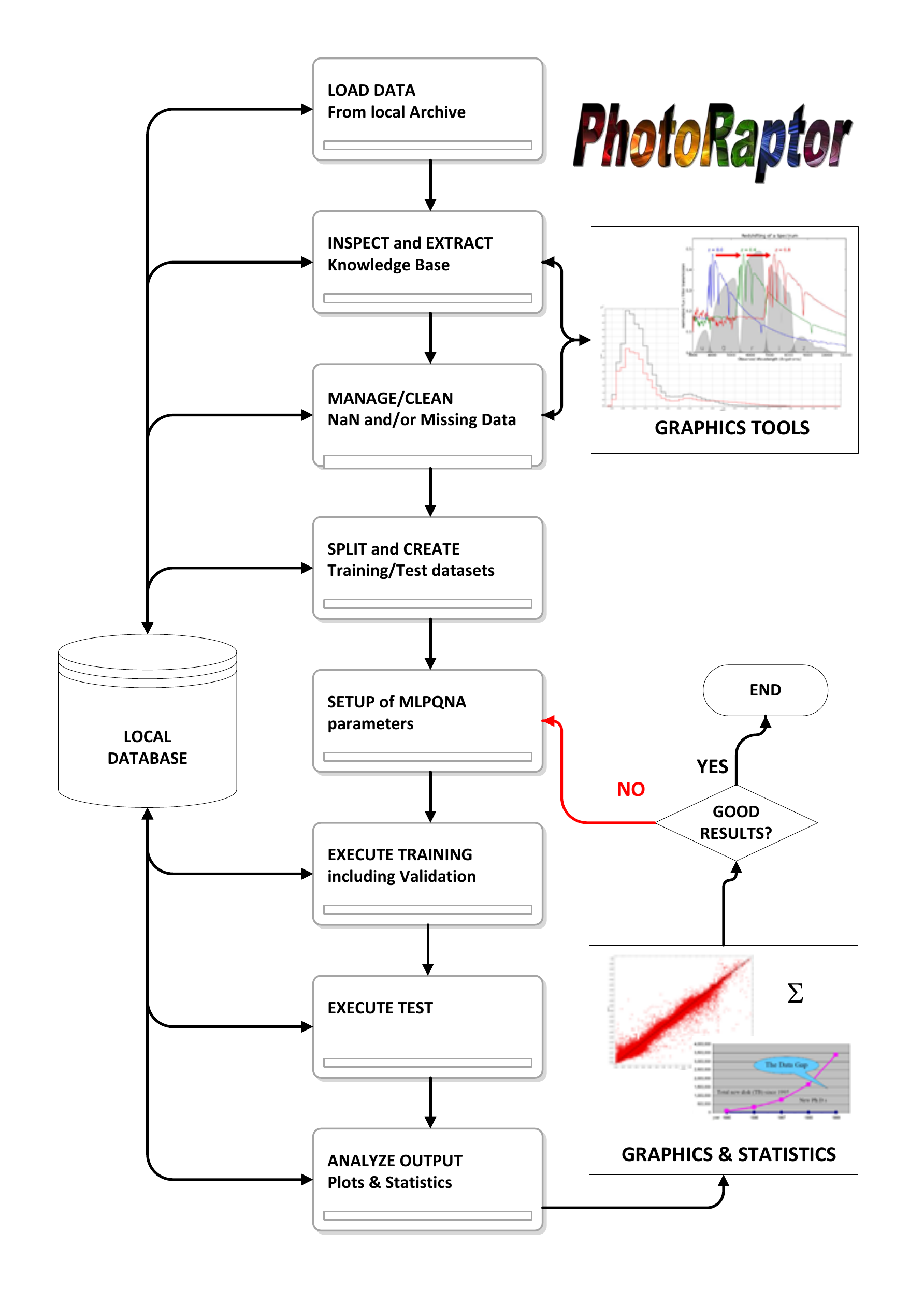}
\caption{The workflow of a generic experiment with PhotoRApToR.}
\label{fig:workflow}
\end{figure*}

In Fig.~\ref{fig:workflow} the layout of a general PhotoRApToR experiment workflow is shown. It is valid for either regression
and classification cases.

%%%%%%%%%%%%%%%%%%%%%%%%%%%%%%%%%%%%%%%%
% CHAPTER 1
\chapter{Installation}
\label{installation}
%%%%%%%%%%%%%%%%%%%%%%%%%%%%%%%%%%%%%%%%

%--------------------------------------
% SECTION 1.1
\section{Download and System Requirements}
\label{releases}
%--------------------------------------

The PhotoRApToR program package is available at the official website (\url{http://dame.dsf.unina.it/dame_photoz.html#photoraptor}).
Here there are different downloadable versions (zipped files) for different OS (Operative System) platforms:

\begin{itemize}
\item $WIN7$: (archive name $PhotoRApToR\_win7.zip$), package for MS Windows $7$, generic platform type;
\item $WIN8$: (archive name $PhotoRApToR\_win8\_64.zip$), package for MS Windows $8$, $64-bit$ platform;
\item $UBUNTU64$: (archive name $PhotoRApToR\_Ubuntu\_64.zip$), package for Linux Ubuntu v12.04, $64-bit$ platform;
\item $SL64$: (archive name $PhotoRApToR\_SL6\_64.zip$), package for Scientific Linux 6, $64-bit$ platform;
\item $MacOSX10.7.5 (Lion)$: (archive name $PhotoRApToR\_MacLion.zip$), package for Mac OS X Lion;
\item $MacOSX10.9.2 (Mavericks)$: (archive name $PhotoRApToR\_MacMavericks.zip$), package for Mac OS X Mavericks;
\end{itemize}

Several other versions could be made available on request.\\

In terms of system requirements, the user must have previously installed and verified the JVM, available at the official site \url{http://www.oracle.com/technetwork/java/index.html}. No other specific requirements are needed.

%--------------------------------------
% SECTION 1.2
\section{Installation Procedure}
\label{procedure}
%--------------------------------------

After having chosen and downloaded the package, the following are the main steps to install it.

\begin{enumerate}
\item unzip the package compressed archive in a user selected relative path, hereinafter called as $<user\_path>$.
\item \textbf{IMPORTANT NOTE}: For the $WIN7$ and $WIN8$ versions we inform that when the program is launched for the first time, there will be automatically created a working directory, named \textit{PhotoRApToR}, in the path \textit{C:}. Here all further experiment executions and results will be stored.
\item Launch the Java executable file \textit{PhotoRApToR} (for its location see the Sec. \ref{verification}).
\end{enumerate}

%--------------------------------------
% SECTION 1.1
\section{Verification}
\label{verification}
%--------------------------------------

After having downloaded and installed the software package on your machine, in the chosen relative path $<user\_path>$ you would have a directory called \textit{PhotoRApToR}. This is the parent directory hosting all files as needed to run the program.
In particular under the parent directory you would be able to find the following directory tree and contents:

\begin{description}
\item[PhotoRApToR jar file] the main program Java executable. This is the file to be launched to run the program.
\item[lib] the directory containing all libraries and executables. \textbf{Don't change or remove its contents}. The included files should be:
 \begin{description}
  \item[.Jar files:] a series of Java executable files. They may change according the specific version installed, depending on your local OS;
  \item[.Exe files:] a series of executable files. They may change according the specific version installed, depending on your local OS;
  \item[.Dll files:] a series of dynamic library files. They may change according the specific version installed, depending on your local OS;
 \end{description}

\item[resources] the directory containing some package internal utilities. \textbf{Don't change or remove its contents}. The included files should be:
 \begin{description}
  \item[.Png files:] a series of png type image files and icons internally used by the program. They may change according the specific version installed, depending on your local OS;
  \item[.Pdf files:] at least the user manual of the program (this document), should be present;
 \end{description}

\item[TEST] the directory containing some additional files. The included files should be:
 \begin{description}
  \item[.Csv files:] a series of CSV type files, useful to be used as examples of data files for experiments during the initial exploration of the program facilities. In particular at least one file useful for regression and one for classification exercises should be present.
 \end{description}
\end{description}

In the case of $WIN7$ and $WIN8$ releases, after first launch of the program, under the path $C:$ you will find a directory \textit{PhotoRApToR}, containing a subdir:

\begin{description}
\item[MyExp] the directory which can be used as destination path for user experiments. At the beginning this directory is empty. Whenever the user runs an experiment, this directory will be populated by sub-directories with name based on date and time of the experiment execution. At the user convenience, all experiment sub-directories can be removed and/or edited without problems.
\end{description}

For the $Ubuntu64$ release this subdir will be located in the same relative destination path of the installation package.

In order to verify the correct execution of the program, we suggest the user to inspect the program parent directory tree to check the presence of all sub-directories and files as described above. Then the user can launch the PhotoRApToR Java executable to see if the program correctly runs.
As first action, we suggest the user to try all menu options, by previously loading at least one data table file as first example.

In case of any wrong behavior or failure, please contact us through the e-mail \textit{helpdame@gmail.com}, by specifying details of the problem and current installed version.

%%%%%%%%%%%%%%%%%%%%%%%%%%%%%%%%%%%%%%%%
% CHAPTER 2
\chapter{The program menu options}
\label{menu}
%%%%%%%%%%%%%%%%%%%%%%%%%%%%%%%%%%%%%%%%

The main window of the PhotoRApToR application (see Fig. \ref{fig:main}) is divided in three parts.\\
The first one is the \textbf{Menu Bar} with a \textbf{Button Bar} below, the second one is the \textbf{Table List} on the left and the third one
is the panel on the right with \textbf{Table Properties}, \textbf{Table Editor} and the \textbf{Split Panel} below.

\begin{figure}[t]
\centering
\includegraphics[width=8cm]{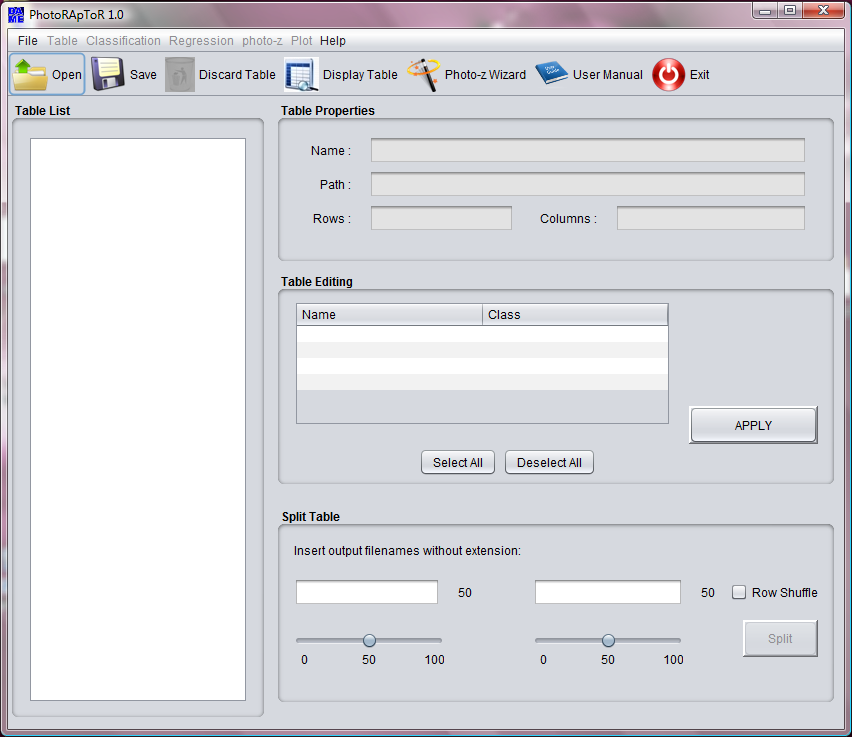}
\caption{The PhotoRApToR main window.}
\label{fig:main}
\end{figure}

Beginnig from the \textit{Menu Bar}, it is possible to decribe all the commands:
\begin{description}

 \item[File] is the menu from which to launch standard commands to open or save files. The following options are shown:
 \begin{description}
  \item[Load Table:] opens a new dialog where it is possible to select table format and file;
  \item[Discard Table:] allows to erase a table item from the \textit{Table List};
  \item[Save Table:] saves the selected table using a Browse Dialog;
  \item[Exit]
 \end{description}

 \item[Table] is a menu containing the commands that allow to see and modify the table properties:
 \begin{description}
  \item[Table Data:] it opens the selected table in a new window;
  \item[Table Metadata:] shows only the column's metadata for the selected table;
  \item[Row Shuffle:] the selected table rows are shuffled and the new table is opened in a new window;
  \item[Not a Number:] it opens a new window for managing the table data in order to remove the Not a Number elements (e.g. values like $-9999$) from the dataset;
 \end{description}

 \item[Classification] is a menu where are selectable different options that allow to execute different experiments:
 \begin{description}
  \item each one of the options \textbf{Train, Test} or \textbf{Run} opens a new window where it is possible to configure and execute experiments using the model MLPQNA as engine;
 \end{description}

 \item[Regression] is a menu with other experimental options:
 \begin{description}
  \item the options \textbf{Train, Test} and \textbf{Run} are the same of those present in the Classification menu;
  \item[Statistics:] in a new window the user can select the Target column and the Output column to generate related statistics;
  \item[Outliers:] it opens a new window where the user can analyze outliers of a loaded data table and generate a dataset without outliers by setting statistical parameters;
 \end{description}

 \item[photo-z] performs experiments dedicated to the evaluation of photometric redshift in a new window. It sets MLPQNA parameters and generates an output table and the related statistics;

 \item[Plot] is the menu that shows three different ways to generate data plots:
 \begin{description}
  \item[Histo Plot:] it opens a new window where to create single or multiple histograms;
  \item[Scatter Plot:] it opens a new window where to create single or multiple scatter plots. It contains several parameters to set up, like for instance the type of line plot or the marker for the data points;
  \item[3D Plot:] after the selection of three table columns and the setup of parameters, a cube plot is generated with also the possibility to change the angle of view;
 \end{description}

 \item[Help] is the menu with manuals and program's credits;
 \begin{description}
  \item[User's Manual:] it opens this document;
  \item[Open Wizard:] it runs the Primer Wizard guided procedure;
  \item[Website:] the program info site (\url{http://dame.dsf.unina.it/dame_photoz.html#photoraptor});
  \item[About:] the application information;
 \end{description}

\end{description}

The \textit{Quick Button Menu bar}, located under the main menu bar, allows a fast access to the main functions of the application.

\begin{description}
 \item[Open] is a button that opens a dialog for the selection of table format and the browsing of files (see Fig. \ref{fig:load});
 \item[Save] opens a dialog to save the tables;
 \item[Discard Table] removes selected data tables from the application table list (not removing the physical table file);
 \item[Display Table] shows the whole table dataset in a new window;
 \item[Photo-z Wizard] starts the Primer Wizard window;
 \item[User Manual] opens this document;
 \item[EXIT] closes the application and exits.
\end{description}

%%%%%%%%%%%%%%%%%%%%%%%%%%%%%%%%%%%%%%%%
% CHAPTER 3
\chapter{Pre-processing}
\label{pre}
%%%%%%%%%%%%%%%%%%%%%%%%%%%%%%%%%%%%%%%%

The evaluation of photo-z is made possible by the existence of a rather complex and not analytically known correlation existing among the fluxes, as measured in broad band photometry, the morphological types of the galaxies, and their distance. The search for such a correlation (a nonlinear mapping between the photometric parameter space and the redshift values) is particularly suited to data mining methods.  For data sets in which accurate and multi-band photometry for a large number of objects is complemented by spectroscopic redshifts, and for a statistically significant sub-sample of the same objects, the \textit{empirical methods} offer greater accuracy. These methods use the sub-sample of the photometric survey with spectroscopically-measured redshifts as a training set to constrain the parameters of a fit mapping the photometric data as redshift estimators.\\

The fundamental premise to use this tool is that the user must preliminarily know how to represent its data. As trivial as it might seem, it is worth to explicitly state that depending on the ML method employed, the user must: \textit{(i)} be conscious of the target of his experiment, such as for instance a regression or classification; and \textit{(ii)} possess a deep knowledge of the characteristics and of the meaning of his data.\\

\begin{figure}[t]
\centering
\includegraphics[height=5cm]{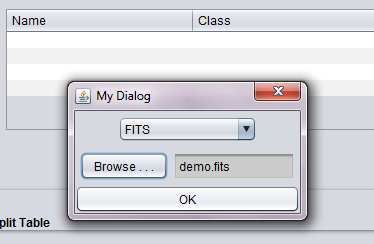}
\caption{The Load Table dialog.}
\label{fig:load}
\end{figure}

The first step is to open a table by selecting the file format and by browsing it through the Load Dialog (Fig. \ref{fig:load}).

In order to reach an intelligible and homogeneous representation of data sets, it is mandatory to preliminarily take care of their internal
format to transform pattern features and force them to assume a uniform representation before submitting them to the training process. In this
respect real working cases might be quite different among themselves.

PhotoRApToR can ingest and/or produce data in any of the following supported formats:
\begin{itemize}
\item	FITS \citep{wells1981}: tabular/image;
\item	ASCII \citep{ansi1977}: ordinary text, i.e. space separated values;
\item	VOTable\footnote{http://www.ivoa.net/documents/VOTable/}: VO compliant XML-based documents;
\item	CSV \citep{repici2010}: Comma Separated Values;
\end{itemize}

In the Load Dialog a drop down menu allows to select the file format. By clicking on the \textbf{Browse} button, it is possible to search the local dataset.

For this description a file named ``\textbf{demo.fits}'' was choosen: every time a new table is loaded, a new item, with its table name, is added to the \textit{Table List}.

%--------------------------------------
% SECTION 3.1
\section{User data handling}
\label{data}
%--------------------------------------

The PhotoRApToR core engine is the MLPQNA neural network. In this respect, before launching any experiment, it may be necessary to manipulate data in order to fulfill the requirements in terms of training and test patterns (data set rows) and features (data set columns) representation as well as contents: \textit{(i)} either training and test data files must contain the same number of input and target columns, in the same order; \textit{(ii)} the target columns must always be the last columns of the data file; \textit{(iii)} the input columns (features) must be limited to the physical parameters, without any other type of additional columns (like column identifiers, object coordinates etc.); \textit{(iv)} all
input data must be numerical values (no categorical entries are allowed).

The application makes available a set of specific options to inspect and modify data file entries.
Selecting one item from the \textit{Table List}, all the table properties are displayed inside the right panel: in particular the table name, its complete path and the number of columns and rows. With a double click on the table name in the Table List or by clicking the \textbf{Display Table} button it is possible to open a new window showing the complete dataset.

\begin{figure}[t]
\centering
\includegraphics[height=8cm]{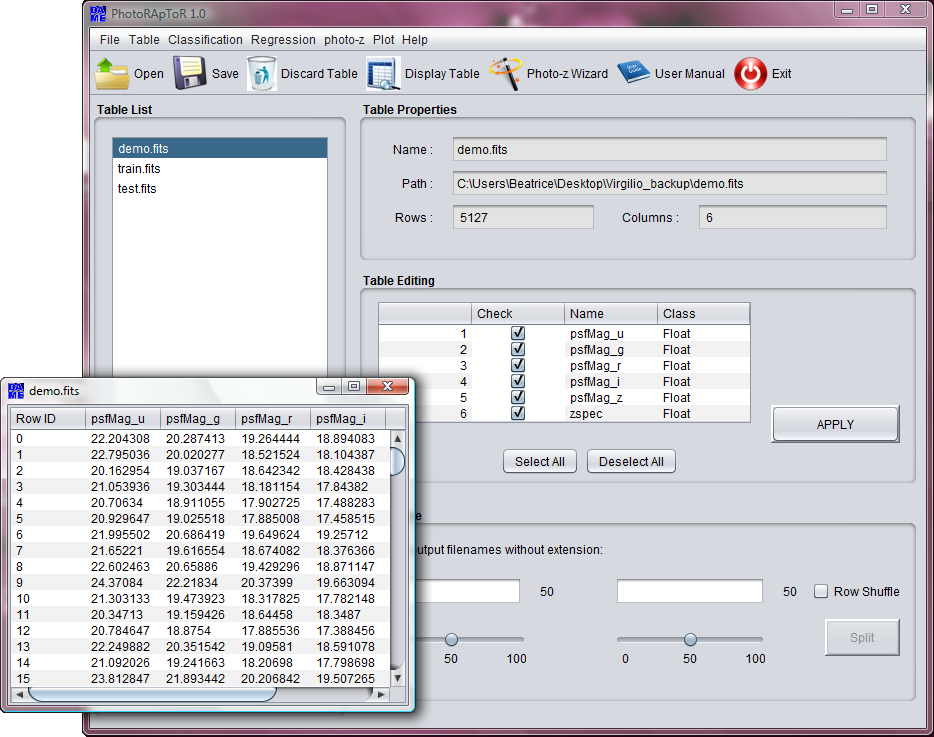}
\caption{Every dataset in the table list can be displayed in a new Table Window.}
\label{fig:display}
\end{figure}

%%%%%%%%%%%%%%%%%%%%%%%
% SECTION 3.1.1
\subsection{Data Feature Selection}
%%%%%%%%%%%%%%%%%%%%%%%%%

A fundamental step for any machine learning experiment is to decide which features to use as input attributes for patterns to be learned. In the specific case of photo-z estimation from a spectroscopic KB, from the available data it is thus necessary to inspect and check which types of flux combinations would be more effective, in terms of available photometry (number and type of fluxes, bands, magnitudes or related colors).

In practical sense, one has to try to maximize the information carried by hidden correlations among different bands, magnitudes and zspec available. In spite of what can be thought, not always the maximum number of available parameters should be used to train a machine learning model. The experience has demonstrated that it is more the quality of data, than the quantity of features and patterns, the crucial key to obtain best prediction results \citep{brescia2013a}. Of course, it depends on how wide is the variety of photometric bands and magnitudes for which a high quality of zspec entries is available in the KB. As usual the cross-matching among different surveys makes available a wider number of
photometric bands, but sometimes drastically reducing the number of objects available for training (i.e. the KB). But if the related photometry quality is sufficiently high, this is the best way to obtain good performances.

In the \textit{Edit Table} panel all column \textit{meta-data} for the selected table are displayed and with them it is possible to generate a subset of the table containing the needed columns only. After the selection of desired columns using the related checkboxes, a table subset is created by clicking the \textbf{Apply} button.\\

\begin{figure}[t]
\centering
\includegraphics[width=\textwidth]{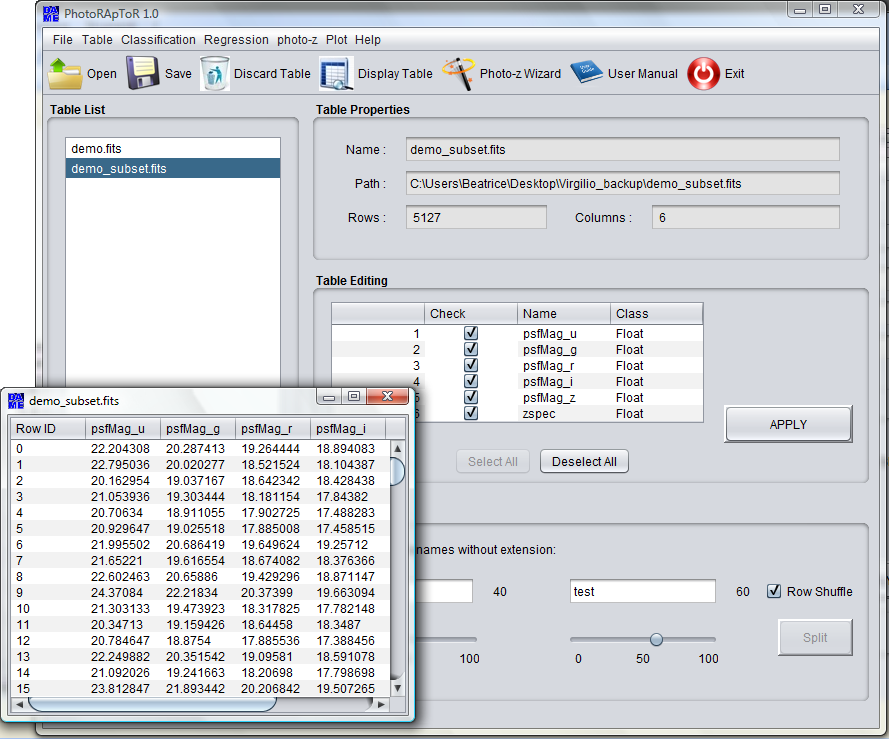}
\caption[Split Tool]{\footnotesize Table subset creation. With the click of the \textbf{Apply} button a new dataset has been generated,
with only the original dataset features in the columns selected in the Edit Table panel.}
\label{fig:subset}
\end{figure}

%%%%%%%%%%%%%%%%%%%%%%%
% SECTION 3.1.2
\subsection{Data Editing}
\label{editing}
%%%%%%%%%%%%%%%%%%%%%%%%%

\begin{figure}[ht!]
\centering
\includegraphics[height=8cm]{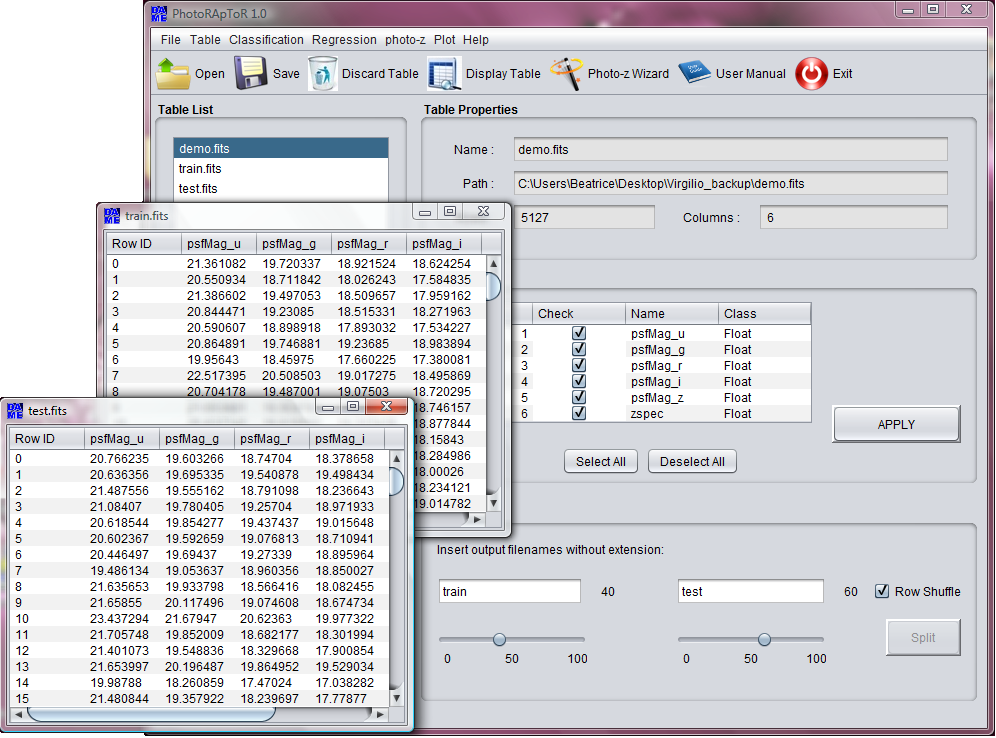}
\caption[Split Tool]{\footnotesize Use of the Split tool. After
selecting the table to be split, two different names are choosen for the files and the sliders are dragged to select a different percentage. By clicking on Split button the two split datasets are generated.}
\label{fig:splitool}
\end{figure}

The random shuffling operation is useful to avoid systematic trends during training and to ensure homogeneity in the distribution of training and test patterns. This last property is, in fact, directly connected to the necessity to split initial data into separated sets, respectively for training and for testing phases. This is a simple action made possible by the \textit{Split} option. When the table is selected in the \textit{Table List}, we must choose two different names for the split files (in this case \textit{train} and \textit{test}) and two different percentages of the original data set\footnote{It is important to observe that for machine learning supervised methods three different subsets for every experiment would be generally required from the available KB: \textit{(i)} (\textit{training set}) to train the method in order to acquire the hidden correlation among the input features; \textit{(ii)} the (\textit{validation set}), used to check and validate the training in particular against the loss of generalization capabilities (a phenomenon also known as overfitting); and \textit{(iii)} the (\textit{test set}), used to evaluate the overall performances of the model (\citealt{brescia2013a}). Within the implemented version of MLPQNA model in the PhotoRApToR application, the validation is embedded into the training phase, by means of the standard leave-one-out k-fold cross validation mechanism (\citealt{geisser1975}).}.\\
By clicking on \textbf{Split} button, the two split datasets are generated. If the selectable checkbox \textbf{Row Shuffle} is selected, the two datasets are also randomly shuffled by rows before to split them, and added to the \textit{Table List} (Fig. \ref{fig:splitool}), ready for the next phase.\\

There is no any analytical rule to decide the percentages of the splitting operation. According the direct experience, an empirical rule of thumb suggests to use $80\%$ and $20\%$ for training and test respectively \citep{kearns1996}. But certainly it depends on the initial amount of available KB. For example also $60\%$ vs $40\%$ and $70\%$ vs $30\%$ could be in principle used in case of large datasets (over ten thousand patterns). It depends also on the quality of available KB. When both photometry and spectroscopy are particularly clean and precise (i.e. with a high S/N), there could also be possible to obtain high performances by training on the half of the KB. The more patterns are available for test, the more precise is the statistical knowledge about experiment performance. But this straightforward rule is valid only in case of a sufficiently large KB, i.e. without affecting the number of patterns necessary to train the network.\\

%%%%%%%%%%%%%%%%%%%%%%%%%%%%%%%%%%%%%%%%
% SECTION 3.2
\section{Not a Number}
\label{NaN}
%%%%%%%%%%%%%%%%%%%%%%%%%%%%%%%%%%%%%%%%

It may be frequent that a data table may have empty entries (sparse matrix) or missing data (lack of observed values for some features in some patterns). Missing values \citep{marlin2008} are frequently identified by special entries in the patterns, like Not-A-Number, out-of-range, negative values in a defined positive numeric field, etc. Missing data is among the most frequent sources of perturbation in the learning process, causing confusion in classification experiments or mismatch in regression problems. This is especially true for astronomy where inaccurate or missing data are not only frequent, but very often cannot be simply neglected, since they carry useful information. To be more specific, missing data in astronomical databases can be of two types:

\noindent Type I: true missing data which were not collected. For instance a given region of the sky or object was not observed in a given photometric band thus leading to a missing information. These missing data may arise also from the simple fact that data, coming from any source and related to a generic experiment, are in most case not expressly collected for DM purposes and, when originally gathered, some features were not considered relevant and thus left unchecked.

\noindent Type II: upper limits or non-detections (i.e. object too faint to be detected in a given band).
In this case the missing datum conveys very useful information which needs to be taken into account into the further analysis. It needs to be noticed, however that, often upper limits are not measured in absence of a detection and therefore this makes these missing data undistinguishable from Type I.\\

In some cases, inaccurate values can be related to systematic data errors, for example due to an hardware setup condition in the data collecting instrument. However, this is usually trivial to be recovered, if the user has knowledge about the collecting device conditions, but in any case a deep care about the presence of inaccurate values should be required whenever a DM process is approached, and a close interaction with a domain expert helps in preventing wrong results in the experiment.

In other words, missing data in a data set might arise from unknown reasons during data collecting process (Type I), but sometimes there are very good reasons for their presence in the data since they result from a particular decision or as specific information about an instance for a subset of patterns (Type II). This fact implies that a special care needs to be put in the analysis of the possible presence (and related causes) of missing values, together with the decision on how to submit these missing data to the ML method, in order to take into account such special cases and to prevent wrong behaviors in the learning process.

In principle, data entries affected by missing attributes, i.e. patterns having fake values for some features, may be used within the KB for a photo-z experiment. In particular they can be used to differentiate the data sets with an incremental quantity of affected patterns, useful to evaluate their noise contribution to the performance of the photo-z estimation after training. Theoretically it should be expected that a greater amount of missing data, evenly distributed in both training and test sets, induces a greater deterioration in the quality of the results. This precious information may be indeed used to assign different indices of quality to the produced photo-z catalogue.

The organization of data sets with different rates of missing data can be performed through PhotoRApToR by means of the following options:\\

\begin{figure}[ht!]
\centering
\includegraphics[width=\textwidth]{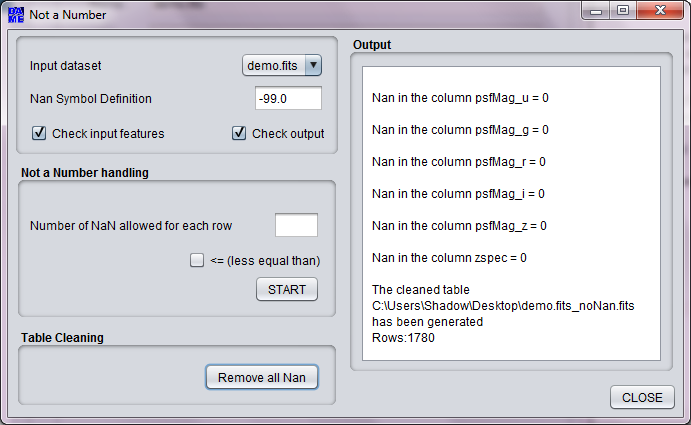}
\caption[Split Tool]{\footnotesize Use of the NaN tool. After the definition of NaN symbol in the input dataset, the user can generate a new dataset only with rows containing NaN elements or one cleaned by NaN. }
\label{fig:NaN}
\end{figure}

A click on \textbf{Table}$>$\textbf{Not a Number} menu item opens a new window. In the upper left panel a drop down menu allows to select the dataset to check and in a text field the user must define the symbol by which missing data are represented in the dataset (i.e. symbols like $-9999$, $-99.0$, $NaN$ and so on). Two checkboxes select which columns will be checked: the input features columns (in this demonstration case the photometric input data), or the output column (in this case, the zspec column).\\

In the panel below it is possible to set the number of NaN for each row, and by clicking on \textbf{START} button a new dataset is generated, built only with rows containing the chosen number of the NaN. If the $\leq$\textit{(less equal than)} checkbox
is selected, the dataset will be generated with all the rows containing less NaN elements than the user has selected.\\
If the user clicks the \textbf{Remove all NaN} button, the tool generates a dataset without all the rows containing NaN elements.\\
In the Output panel on the right side of Fig.~\ref{fig:NaN} the number of NaN for each column and for each row of the input dataset and the path of the output file are reported.

%%%%%%%%%%%%%%%%%%%%%%%%%%%%%%%%%%%%%%%%
% CHAPTER 4
\chapter{Photometric redshift estimation}
\label{photoz}
%%%%%%%%%%%%%%%%%%%%%%%%%%%%%%%%%%%%%%%%

\begin{figure}[t!]
\centering
\includegraphics[height=10cm]{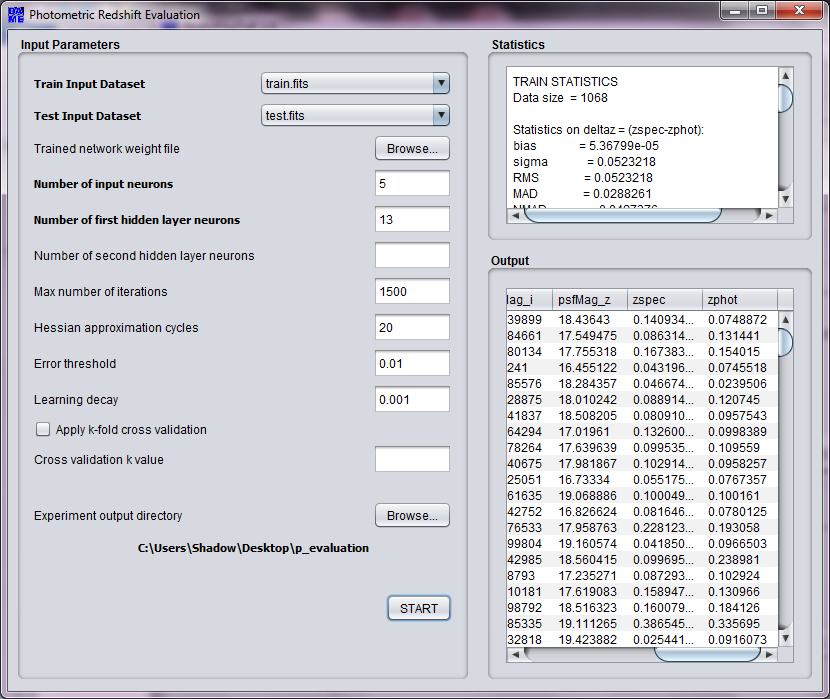}
\caption[Photometric redshift evaluation window]{\footnotesize Photometric redshift evaluation window. On the left there are the fields for setting parameters. On the right there are two panels. When the experiment is complete, in the upper panel the regression train and test statistics is displayed. In the lower panel the final table with the photo-z column is reported.}
\label{fig:photoevaluation}
\end{figure}

A click on the \textbf{photo-z} button opens a new window (Fig. \ref{fig:photoevaluation}).\\
After having prepared the KB (see Chap.~\ref{pre}), the user would have two subset tables ready to be submitted for a photo-z experiment. By looking at the Fig.~\ref{fig:workflow} the experiment consists of a pre-determined sequence of steps, for instance \textit{(i)} Training and validation of the model network; \textit{(ii)} blind Test of the trained model network; \textit{(iii)} statistical and visual inspection of results; and \textit{(iv)} Run, i.e. the execution of a well trained, validated and tested network on new data samples.

We outline that for the first two steps, the basic rule is to use different data subsets. In general all empirical photo-z methods may suffer of a poor capability to extrapolate outside the range of distributions imposed by the parameter space and photometric flux limits used for the training. In other words, outside the limits of magnitudes and spectroscopic redshift (zspec) used in the training set, these methods do not ensure optimal performances. But, within the ranges of the training parameter space, the empirical models are able to overtake fitting models, essentially because they do not
make any a priori assumption on the physical properties of objects\footnote{priors like SFR, IMF, metallicity, age etc.}.
In order to remain in a safe condition, the user must perform a selection of test data according to the training sample photometric and spectroscopic limits.

Therefore, none of the objects included in the training sample should be included also in the test sample. Moreover only the data set used for the test has to be used to generate performance statistics. In other words the test must be blind, i.e. containing only objects never submitted to the network before.

For what the training is concerned, this phase embeds two processing steps, for instance training of the MLPQNA model network and training validation. It is in fact quite frequent for machine learning models to suffer of an \textit{overfitting} on training data, affecting and badly conditioning the training performances. The problem arises from the paradigm of supervised machine learning itself. Any ML model is trained on a set of training data in order to become able to predict new data points. Therefore its goal is not just to maximize its accuracy on training data, but mainly its predictive accuracy on new data instances. Indeed, the more hard and computationally stiff is the model setup during
training, the higher would be the risk to fit the noise and other peculiarities of the training sample in the new data \citep{dietterich1995}. The technique known as \textit{cross validation} does not suffer of such drawback; it can avoid overfitting on data and is able to improve the generalization performance of the ML model.

Therefore in the PhotoRApToR application, the validation can be implicitly performed during training, by enabling at the setup step the standard leave-one-out k-fold cross validation mechanism \citep{geisser1975}. The automatized process of the cross-validation is done by performing k different training runs with the following procedure: (i) splitting of the training set into k random subsets, each one composed by a percentage of the data set (depending on the k choice); (ii) at each run we applied the rest of the data set for training and the excluded percentage for validation. The k-fold cross validation is able to avoid overfitting on the training set, although with an increase of the execution time estimable around $k-1$ times the total number of runs.

The photo-z experiment setup sets the MLPQNA input parameters necessary to run a regression train + test experiment, so that it generates an output table where last column is the estimated photometric redshift. Two drop-down menu allow to select the TRAIN dataset and the TEST one (\textbf{this parameter is a field required}).\\
The other parameters are involved in the regression training only, because for the Test phase only the test input dataset is required and remaining parameters are derived by the internal model configuration as frozen at the end of training.\\

\noindent We can group the MLPQNA model training parameters into three subsets:

\begin{itemize}
\item \textbf{network topology}: all parameters related to the MLP network architecture;
\begin{itemize}
\item \textit{input neurons}: the number of neurons at input layer. In terms of input data set it corresponds to the number of columns of the data table, (also named as input features of the data sample, i.e. magnitudes/colors composing the photometric information of each object in the data), except for the target column (i.e. the spectroscopic redshift), which is related to the single output neuron of the regression network.(\textbf{this parameter is a field required});
\item \textit{first hidden layer neurons}: the number of neurons composing the first layer after the input. As a rule of thumb, it is reasonable to set this number to $2N+1$, where N is the number of input neurons (\textbf{this parameter is a field required});
\item \textit{second hidden layer neurons}: this is an optional parameter. Although not required in normal conditions, as stated by the known universal approximation theorem \citep{cybenko1989}, problems having a very high complexity of its parameter space, i.e. with a large amount of distribution irregularities, are better treated by what was defined as \textit{deep} networks, i.e. networks with more than one computational (hidden) layer \citep{bengio2007}. As a rule of thumb, it is reasonable to set this number to $N-1$, where $N$ is the number of input neurons;
\item \textit{trained network weights}: this parameter is related to the file containing the matrix of weights (internal connections among neurons). A weight matrix exists only after having performed one training session at least. So far this parameter is left empty at the beginning of any experiment. But, for all other use cases (Test, Run) it is required to load an already trained network. However this parameter could also be used to perform further training cycles for an already trained network (i.e. enhanced training);
\end{itemize}
\item \textbf{learning rule setup}: all parameters related to the QNA learning rule;
\begin{itemize}
\item \textit{Max number of iterations}: the maximum number of iterations at each Hessian approximation cycle. Typical range for such value is $[1000, 30000]$, depending on the requested precision. It can affect the computing time of the training;
\item \textit{Hessian approximation cycles}: number of approximation cycles searching for the best value close to the Hessian of the error. If set to zero, the max number of iterations will be used for a single cycle. At each cycle the algorithm performs a series of iterations along the direction of the minimum error gradient, trying to approximate the Hessian value. A reasonable range could be $[20, 60]$, depending on the final precision required. If set to a high value, it is recommended to enable the cross validation option (see below), to prevent overfitting occurrence;
\item \textit{Training error threshold}: This is the stopping criteria of the algorithm. It is the training error threshold (a value of $0.01$ is typical for photo-z experiments);
\item \textit{Learning decay}: this value determines the analytical \textit{stiffness} of the approximation process. It affects the expression of weight updating law, by adding the term $decay*||network weights||^2$. Its range goes from a minimum value of $0.001$ (very low stiffness) up to $100$ (very high stiffness). If set to a high value, it is recommended to enable the cross validation option (see below), to prevent overfitting occurrence;
\end{itemize}
\item \textbf{validation setup}: all parameters related to the optional training validation process;
\begin{itemize}
\item \textit{Cross validation k value}: when the cross validation is enabled, this value is related to the automatic procedure that splits in different subsets the training data set, by applying a k-step cycle in which the training error is evaluated and its performances are validated. A reasonable value could be $5$ or $10$, depending on the amount of training data used. We remind that this value may highly affect the computing time of the experiment.
\end{itemize}
\end{itemize}

Finally \textbf{Experiment output directory} is the parent directory of the output for the experiments and it was called \textbf{p\_evaluation} in the example shown in Fig.~\ref{fig:photoevaluation}.

%%%%%%%%%%%%%%%%%%%%%%%%%%%%%%%%%%%%%%%%
% SECTION 4.1
\section{The training error and decay factor}
\label{decay}
%%%%%%%%%%%%%%%%%%%%%%%%%%%%%%%%%%%%%%%%

The error calculated during training by the MLPQNA is evaluated for all the presented input patterns between their known target and the calculated output of the model. The error function in the regression case is based on the Least Mean Square (LSE) + Tychonov regularization \citep{groetsch1984}. This function is defined as follows:

\begin{equation}
E = \frac{\sum^N_{i=1} (y_i-t_i)^2}{2} + \frac{d||W||^2}{2} \nonumber
\label{eqn:error}
 \end{equation}

where $N$ is the number of input patterns, $y$ and $t$ are the network output and the pattern target respectively, $d$ is the decay input parameter and $W$ the network weight matrix.\\

Regularization of the weight decay is the most important issue within the model mechanisms. When the regularization factor is accurately chosen, then generalization error of the trained neural network can be improved, and training can be accelerated. If the best decay regularization parameter $d$ is unknown, it could be experimented the values within the range from $0.001$ (weak regularization) up to $100$ (very strong regularization). In order to achieve the weight decay rule, we minimize more complex merit function:

\begin{equation}
f = E + \frac{dS}{2} \nonumber
\label{eqn:merit}
 \end{equation}

Here $E$ is the training set error, $S$ is the sum of squares of network weights, and decay coefficient $d$ controls the amount of smoothing applied to the network. Optimization is performed from the initial point and until the successful stopping of the optimizer has been reached.\\
Searching for the best decay value is a typical trial-and-error procedure. It is usually performed by training the network with different values of the decay parameter $d$, from the lower value (no regularization) to the infinite value (strongest regularization). By inspecting statistical results at each stage of the procedure (optimal decay can be selected by using test set or cross-validation, and in the latter case all dataset can be used for training), it can be seen the control tendency to overfit by continuously changing the decay factor. A zero decay corresponds to overfitted network. Infinitely large decay gives us an underfitted network. Between these extreme values there is a range of networks which reproduce dataset with different degrees of precision and smoothness.

%%%%%%%%%%%%%%%%%%%%%%%%%%%%%%%%%%%%%%%
% SECTION 4.2
\section{Primer Wizard}
\label{wiz}
%%%%%%%%%%%%%%%%%%%%%%%%%%%%%%%%%%%%%%%%

When the program is launched, in addition to the main program window, also a tutorial wizard is started, called \emph{Photo-z Primer Wizard}.
\begin{enumerate}
 \item  The first dialog explains scientifical applications of the program and gives the possibility to skip the tutorial by switching to the application main window.
\item In the second dialog it is possible to open table data (selectable choices: ASCII, FitsTable, CSV, VoTable) (cf. Fig.~\ref{fig:wizard2}).

\begin{figure}[h!]
\centering
\includegraphics[height=5cm]{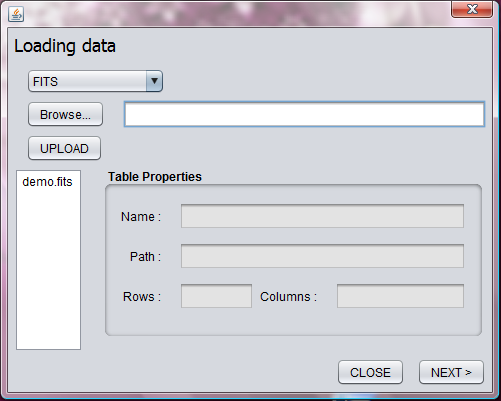}
\caption[wizard1]{\footnotesize Primer Wizard second dialog window.}
\label{fig:wizard2}
\end{figure}

\item In the third dialog it is possible to manipulate metadata to select only needed columns by a checkbox;
 \item In the fourth dialog we can separate our data into two files (\textit{train} and \textit{test}) using the Split function.
\item In the fifth dialog the experiment setup begins (Fig.~\ref{fig:wizard4}). Here it is necessary to insert the parameters to setup the experiment and select the output folder. Then a click on the START button executes the experiment and a status message shows that the experiment is running.

\begin{figure}[h!]
\centering
\includegraphics[height=5cm]{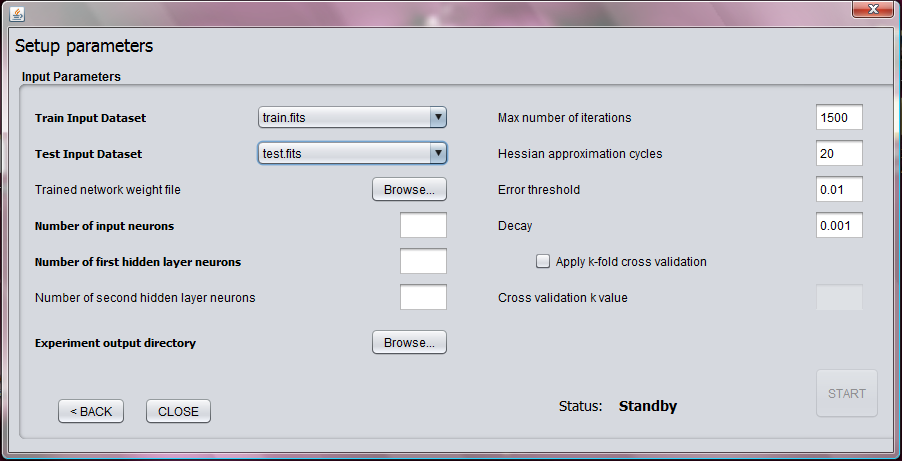}
\caption[wizard1]{\footnotesize Primer Wizard fifth dialog window.}
\label{fig:wizard4}
\end{figure}

\item At the end of the experiment, the final dialog automatically is opened (Fig.~\ref{fig:wizard5}). Here are displayed the output table on the right and the statistical report on the left. By clicking on the Scatter Plot button, in a different dialog the scatter plot zphot/zspec is shown.

\begin{figure}[h!]
\centering
\includegraphics[height=5cm]{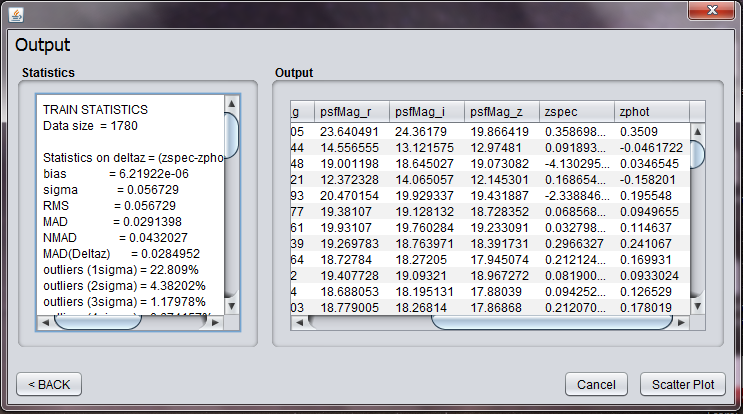}
\caption[wizard1]{\footnotesize Primer Wizard final dialog window.}
\label{fig:wizard5}
\end{figure}

\end{enumerate}

%%%%%%%%%%%%%%%%%%%%%%%%%%%%%%%%%%%%%%%%
% CHAPTER 5
\chapter{Other functionalities}
\label{other}
%%%%%%%%%%%%%%%%%%%%%%%%%%%%%%%%%%%%%%%%

To complete the description of the resources made available by the PhotoRApToR application, we outline that besides photometric redshift estimation, as a specialized kind of regression experiments, the user has the possibility to perform generic regression as well as multi-class classification experiments.

For a generic regression problem, all the above functionalities described in the case of photo-z, remain still valid, with the only straightforward exception for the statistics produced, which is generated for generic quantities formulated below.

\begin{eqnarray}
 \Delta out &=& target - output \nonumber\\
 \Delta out_{norm} &=& \frac{target - output}{1 + target} \nonumber
\end{eqnarray}

%%%%%%%%%%%%%%%%%%%%%%%%%%%%%%%%%%%%%%%%
% SECTION 5.1
\section{Regression}
\label{reg}
\begin{figure}[t]
\centering
\includegraphics[height=8cm]{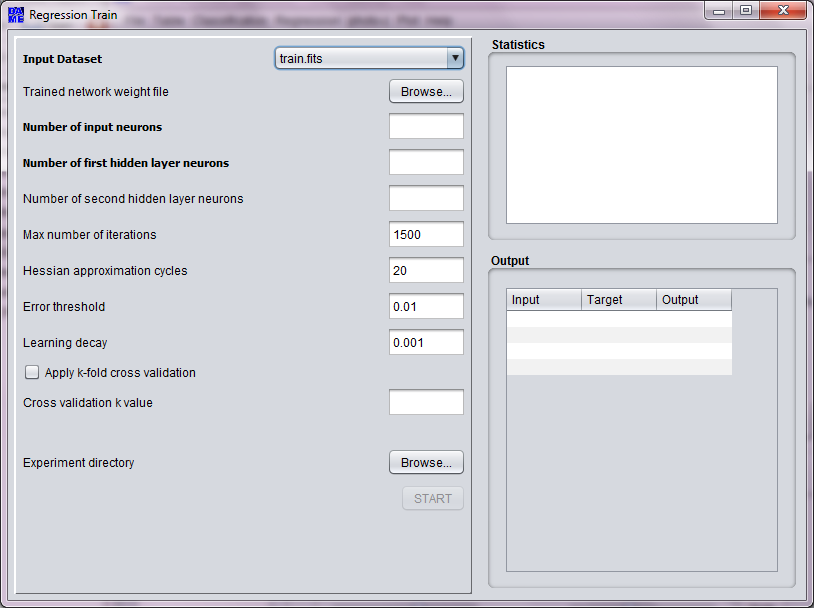}
\caption[Regression Train window]{\footnotesize Regression Train window. On the left there are the fields for setting parameters. On the right there are two panels. When the experiment is complete, in the upper panel the regression train statistics is displayed. In the lower panel the final table with the photo-z column is reported.}
\label{fig:regression}
\end{figure}

%%%%%%%%%%%%%%%%%%%%%%%%%%%%%%%%%%%%%%%%
A click on \textbf{Regression}$>$\textbf{Train} menu item opens a new window (Fig. \ref{fig:regression})
where it is necessary to set MLPQNA's input parameters.
\begin{itemize}
 \item A drop-down menu allows to select the input file; (\textbf{this parameter is a field required})
 \item if we had already done the training phase, it is possible to use the \textbf{trained weight file};
 \item the \textbf{Number of input neurons} is the number of input dataset columns (except for the target column); (\textbf{this parameter is a field required})
 \item the \textbf{Number of first hidden layer neurons} is the number of neurons of the first hidden layer of the network; (\textbf{this parameter is a field required})
 \item the \textbf{Number of second hidden layer neurons}, as a suggestion this number should be smaller than the previous layer. By default the second hidden layer is empty (not used);
 \item \textbf{Max number of iteration} is one of the internal model parameters. It indicates the number of algorithm iterations and it is one of the stopping criteria. By default this value is set to 1500;
 \item \textbf{Hessian approximation cycles} indicates the number of restarts for each approximation step of the Hessian inverse matrix. By default this value is set to 20;
 \item \textbf{Error threshold} indicates the minimum network error at each iteration step (see Sect.~\ref{decay} for details). Except for problems which are particularly difficult to solve, in which a value of 0.0001 should be used, a value of 0.01 is usually considered sufficient. By default this value is therefore set to 0.01;
 \item \textbf{Decay} indicates the weight regularization decay. If accurately chosen, this parameter leads to an important improvements of the generalization error of the trained neural network and implies an acceleration of training (see Sect.~\ref{decay} for details). By default the value is set to 0.001;
 \item \textbf{Cross validation} is based on an automatic procedure that splits in different subsets the training dataset, by applying a k-step cycle in which the training error is evaluated and its performances are validated. By default the k value is set to 10;
 \item finally \textbf{Experiment output directory} is the parent directory hosting the output for the experiments.
\end{itemize}

After the parameter setup, a click on \textit{START} button executes the MLPQNA regression experiment and the resulting output is displayed in the main panel on the left. After the experiment, also the statistics is generated with a specific algorithm and the result is presented in the text panel on top of the panel (Fig.~\ref{fig:regression}).\\

%%%%%%%%%%%%%%%%%%%%%%%%%%%%%%%%%%%%%%%%
% SECTION 5.2
\section{Classification}
\label{class}
%%%%%%%%%%%%%%%%%%%%%%%%%%%%%%%%%%%%%%%%

In the case of the multi-class classification, the above considerations and options remain still valid with only some differences, described in what follows.

During the training setup, there are two specific parameters, not involved in regression problems:

\begin{itemize}
\item \textit{Output neurons}: the number of neurons of the output layer. Forced to $1$ in the obvious case of regression experiments, this parameter corresponds here to the number of different classes present in the training sample. It is required that the class identifiers should have a binary coding format for labels (for example a three-class problem is represented by three columns, labeled as, respectively, $1,0,0$, $0,1,0$ and
$0,0,1$);
\item \textit{Cross entropy}: this optional parameter, if enabled, replaces the standard training error evaluation (for instance the MSE between output and target values). Its meaning is discussed below.
\end{itemize}

\begin{figure}[t]
\centering
\includegraphics[height=8cm]{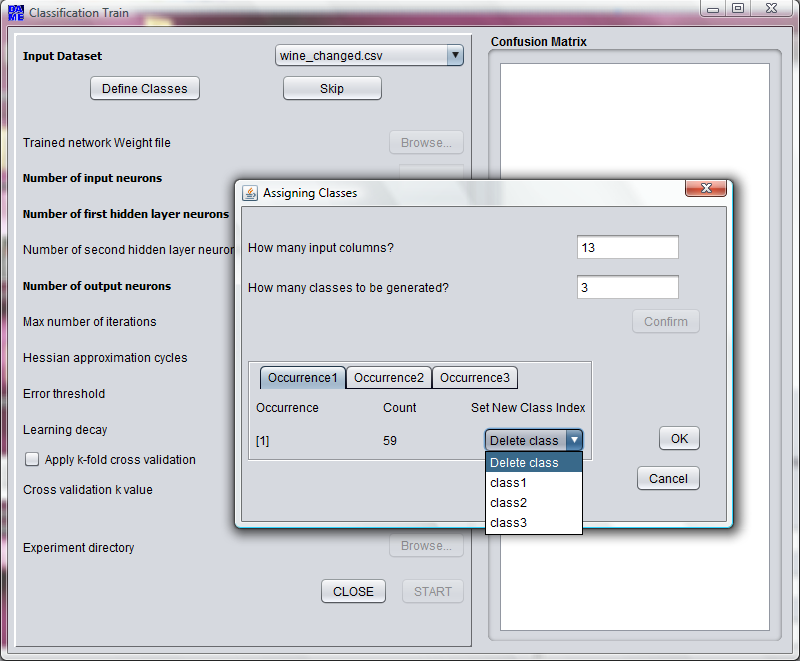}
\caption[Classification Train window]{\footnotesize Classification Train window. On the left there are the fields for setting parameters. On the right panel the Confusion Matrix is displayed.}
\label{fig:classification}
\end{figure}

The option \textbf{Classification}$>$\textbf{Train} opens a new window (Fig. \ref{fig:classification})
for the MLPQNA's parameters setting.
\begin{itemize}
 \item A drop-down menu allows to select the input file; (\textbf{this parameter is a field required});
 \item Two buttons, one to \textbf{Define Classes} and the other to \textbf{Skip} this operation and set the other parameters;
 \item if we had already done the training phase, it is possible to use the \textbf{trained weight file};
 \item \textbf{Number of input neurons:} as specified for regression; (\textbf{this parameter is a field required});
 \item \textbf{Number of first hidden layer neurons:} as specified for regression; (\textbf{this parameter is a field required});
 \item \textbf{Number of second hidden layer neurons:} as specified for regression;
 \item the \textbf{Number of output neurons} is the number of neurons in the output layer of the network. It must correspond to the number of target columns in the input file, represented in binary code; (\textbf{this parameter is a field required});
 \item \textbf{Max number of iteration:} as specified for regression;
 \item \textbf{Hessian approximation cycles:} as specified for regression;
 \item \textbf{Error threshold:} as specified for regression;
 \item \textbf{Decay:} as specified for regression;
 \item \textbf{Cross validation:} as specified for regression;
 \item finally \textbf{Experiment output directory} is the parent directory of the output for the experiments.
\end{itemize}

The \textbf{Define Classes} button open a dialog where the user must set the number of input features and the number of classes that will have as output, codified in a binary representation. By clicking on the \textbf{Confirm} button, the tool will check all the occurrences that will be labeled in binary format to become a correct class identifier for the MLPQNA. The user can choose which binary label should be associated to every occurrence or which occurrence should be deleted. By clicking on the \textbf{OK} button, it is created the modified table and closed the dialog, coming back to the Classification setup window, showing the field \textbf{\textit{Table used}} with the name of the table that will be used during the classification experiment (if the user clicks on the \textbf{Skip} button the input table's name will be in this field).\\

A click on the \textit{START} button executes the MLPQNA classification experiment and the resulting output is displayed
in the main panel on the left. The text panel above the \textit{Confusion Matrix} is reported.

By clicking on \textbf{Test} or \textbf{Run} options of \textit{Regression} and \textit{Classification} menu items, a window similar to those described for the \textbf{Train} case is opened.\\

The Cross Entropy (CE) error function was introduced to address classification problem evaluation in a consistent statistical fashion \citep{rubin2004}. The CE method consists of two phases: \textit{(i)} generate a random data sample (trajectories, vectors, etc.) according to a specified mechanism; \textit{(ii)} update the parameters of the random mechanism based on the data to produce a \textit{better} sample in the next iteration.

In practice a data model is created based on the training set, and its CE is measured on a test set to assess how accurate the model is in predicting the test data. The method compares indeed two probability distributions, $p$ the true distribution of data in any corpus, and $q$ which is the distribution of data as predicted by the model. Since the true distribution is unknown, the CE cannot be directly calculated, while an estimate of CE is obtained using the following expression:

$$H\left(T,q \right)=- \sum_{i=1}^{N} \frac{1}{N} log_2 q\left(x_i \right)$$

where $T$ is the chosen training set, corresponding to the above mentioned true distribution $p$, $N$ is the number of objects in the test set, and $q\left( x \right)$ is the probability of the event $x$ estimated from the training set.

%%%%%%%%%%%%%%%%%%%%%%%%%%%%%%%%%%%%%%%%
% CHAPTER 6
\chapter{Post-Processing}
\label{post}
%%%%%%%%%%%%%%%%%%%%%%%%%%%%%%%%%%%%%%%%

After having successfully terminated a training session, the model will produce (among several output files) a final network weight matrix (file by default called \textit{trainedWeights.txt}) and the network configuration setup (file by default called \textit{frozen\_train\_net.txt}), which can be used during next experiment steps (Test and Run use cases), together with their respective input data sets.

%%%%%%%%%%%%%%%%%%%%%%%%%%%%%%%%%%%%%%%%
% SECTION 6.1
\section{Statistics}
\label{stat}
%%%%%%%%%%%%%%%%%%%%%%%%%%%%%%%%%%%%%%%%

As already underlined, concerning the performance evaluation in terms of photometric redshift reconstruction, all statistical results reported throughout this paper are referred to test data sets only. In fact, it is good practice to evaluate the results on data (i.e. the test set) which have never been presented to the network during the training and/or validation phases. The usage of \textit{test plus training} data might introduce an obvious positive systematic bias which could mask reality.

More in general, empirical methods, such as MLPQNA, have the advantage that the training set is made up of real sky objects. Hence they do not suffer from the uncertainty of having accurate templates. In this sense any empirical method intrinsically includes effects such as the filter band-pass and flux calibrations. In fact, as deeply discussed by \cite{collister2004}, one of the main drawbacks of these methods is the difficulty in extrapolating to regions of the input parameter space that are not covered and well sampled by the training data. Therefore the efficiency of empirical methods degrades for objects at fainter magnitudes than those included in the training set, as this would require an extrapolation capability on data having properties, such as redshift and photometry, not included in the learned sample. In fact, another strong requirement of such methods is that the training set must be large enough to cover properly the parameter space in terms of fluxes, colors, magnitudes, object types and redshift. In this case the calibrations and corresponding uncertainties are well known and only limited extrapolations beyond the observed locus in color-magnitude space are required. In conclusion, under the conditions described above about the consistency of the training set, a
realistic way to measure photometric uncertainties is to compare the photometric redshifts estimation with spectroscopic measures in the test samples.

The obtained results of the individual experiments have to be evaluated in a consistent and objective manner through an homogeneous set of statistical indicators. Within PhotoRApToR we use a specific algorithm to generate statistics.

For each experiment, given a list of $N$ blind test samples for \textbf{$z_{spec}$} and \textbf{$z_{phot}$}, we define:
\begin{eqnarray}
 \Delta z &=& z_{spec} - z_{phot} \nonumber\\
 \Delta z_{norm} &=& \frac{z_{spec} - z_{phot}}{1 + z_{spec}} \nonumber
\end{eqnarray}

where $\Delta z_{norm}$ is the normalized $\Delta z$. By indicating with $x$ either $\Delta z$ or $\Delta z_{norm}$, we calculate the following statistical indicators:
\begin{eqnarray}
 bias(x) &=& \frac{\sum^N_{i=1} x_i}{N} \nonumber \\
 \sigma (x) &=& \sqrt{\frac{\sum^N_{i=1} \left[x_i - \left(\frac{\sum^N_{i=1} x_i}{N}\right)\right]^2}{N}} \nonumber \\
 RMS(x) &=& \sqrt{\frac{\sum^N_{i=1} x_i^2}{N}} \nonumber \\
 MAD (x) &=& Median (\mid x \mid) \nonumber \\
 NMAD (x) &=& 1.4826 \times Median (\mid x \mid) \nonumber
\end{eqnarray}

There is also a relation between the Root Mean Square (RMS) and the Standard Deviation $\sigma$: $RMS = \sqrt{mean^2 + \sigma^2}$, but $\sigma^2$ is the \textit{variance}, so we have $RMS = \sqrt{mean^2 + variance}$. Therefore, for a direct comparison of results, in terms of distance of $m\sigma$ ($m = 1, 2,...$) from the distribution of $\Delta z$, it is much more precise to use the Standard Deviation as main indicator, rather than the simple RMS.

There is often a confusion about the relation between photometric and spectroscopic redshifts used to apply the statistical indicators. For instance, the performance could be very different if the simple $\Delta z$ is used instead of the $\Delta z_{norm}$. The idea is that the $\Delta z$ cannot represent the best choice in the specific case of photometric redshift prediction.

The velocity dispersion error, intrinsically present within the photometric estimation, is not uniform in a wide spectroscopic sample, and the related statistics is not able to give a consistent estimation at all ranges of redshift. On the contrary, the normalized term $\Delta z_{norm}$ introduces a more uniform information, correlating in a more correct way the variation of photometric estimation, thus permitting a more consistent statistical evaluation at all ranges of spectroscopic redshift.

More in detail:
\begin{eqnarray}
 z &=& \frac{\Delta \lambda}{\lambda} = \frac{\lambda_{obs} - \lambda_{emit}}{\lambda_{emit}} = \nonumber\\
 &=& \frac{\lambda_{obs}}{\lambda_{emit}} - 1  \nonumber \\
 => 1 &+& z = \frac{\lambda_{obs}}{\lambda_{emit}} \nonumber
 \label{eqn:redshiftdef}
\end{eqnarray}

So, differentiating the Eq.~\ref{eqn:redshiftdef}:
\begin{eqnarray}
 dz &=& d \left(\frac{\lambda_{obs} - \lambda_{emit}}{\lambda_{emit}}\right) = \frac{d \lambda_{obs}}{\lambda_{emit}} = \nonumber \\
 &=& \frac{d\lambda_{obs}}{\lambda_{emit}} \frac{\lambda_{obs}}{\lambda_{obs}} = \frac{d \lambda_{obs}}{\lambda_{obs}} (1+z)
\end{eqnarray}

We then obtain:
\begin{equation}
 \frac{dz}{1+z} = \frac{d \lambda_{obs}}{\lambda_{obs}}
\label{eqn:finaldiffer}
 \end{equation}

The term on the right of the Eq.~\ref{eqn:finaldiffer} is exactly the variation between photometric and spectroscopic observed redshift, which is the main focus of the photometric redshift estimation for empirical models which learn its prediction based on the spectroscopic information. This result is invariant to the redshift range considered. In conclusion the term $\frac{dz}{1+z}$ is the best choice on which to apply the statistical operators.

All the described statistical indicators are provided as output of any photo-z estimation test and stored in a dedicated file (by default named as \textit{test\_statistics.txt}). For completeness we also provide a similar statistics file as output of any training session. But its use is suggested only as a quick comparison between training and test, just in order to verify the absence of any overfitting occurrence.\\

\begin{figure}[t]
\centering
\includegraphics[height=8cm]{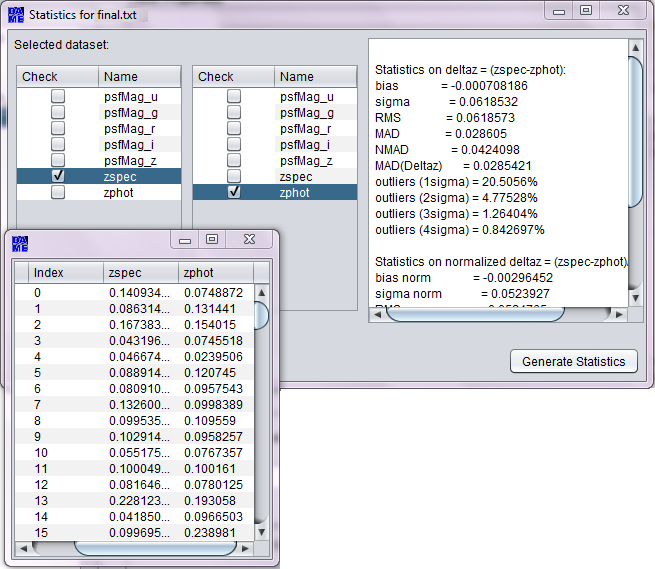}
\caption[Regression statistics window]{\footnotesize Statistic window.}
\label{fig:statistic}
\end{figure}

By clicking on \textbf{Regression}$>$\textbf{Statistic} a new window is opened where it is possible to generate statistics using the described
statistical indicators on local dataset previously loaded in the \textit{Table List} (Fig.~\ref{fig:statistic}).\\
The name of the selected dataset is shown for two lists of its column items, where the user must select one item from the first one and another from the second one. After a click on \textbf{Generate Statistics} button, in the panel on the right the statistical indicators, calculated on the two selected features, will be displayed.

%%%%%%%%%%%%%%%%%%%%%%%%%%%
% SECTION 6.1.1
\subsection{Confusion Matrix}
\label{confmat}
%%%%%%%%%%%%%%%%%%%%%%%%%%%

Another difference in respect of regression experiments is of course the statistics produced to evaluate the results outcoming from a classification experiment. In this case, at the base of the statistical indicators adopted, there is the commonly known confusion matrix, which can be calculated to easily visualize the classification performance \citep{provost1998}: each column of the matrix represents the instances in a predicted class, while each row represents the instances in an actual class. One benefit of a confusion matrix is the simple way to see if the system is mixing different classes.

More specifically, for a generic two-class confusion matrix,

\begin{eqnarray}
  \begin{array}{c|ccc}
            &            & $OUTPUT$\\ \hline
            &    -       &$Class A$      & $Class B$ \\
    $TARGET$  &$Class A$    & N_{AA}              & N_{AB} \\
            &$Class B$    & N_{BA}              & N_{BB} \\
  \end{array} \nonumber
  \end{eqnarray}

we then use its entries to define the following statistical quantities:

\begin{itemize}
\item \underline{total efficiency}: $te$. Defined as the ratio between the number of correctly classified objects and the total number of objects in the data set. In our confusion matrix example it would be:
    \begin{eqnarray}
    te=\frac{N_{AA} + N_{BB}}{N_{AA} + N_{AB} + N_{BA} + N_{BB}} \nonumber
    \end{eqnarray}
\item \underline{purity of a class}: $pcN$. Defined as the ratio between the number of correctly classified objects of a class and the number of objects classified in that class. In our confusion matrix example it would be:
    \begin{eqnarray}
    pcA=\frac{N_{AA}}{N_{AA}+N_{BA}} \nonumber
    \end{eqnarray}
    \begin{eqnarray}
    pcB=\frac{N_{BB}}{N_{AB}+N_{BB}} \nonumber
    \end{eqnarray}
\item \underline{completeness of a class}: $cmpN$. Defined as the ratio between the number of correctly classified objects in that class and the total number of objects of that class in the data set. In our confusion matrix example it would be:
    \begin{eqnarray}
    cmpA=\frac{N_{AA}}{N_{AA}+N_{AB}} \nonumber
    \end{eqnarray}
    \begin{eqnarray}
    cmpB=\frac{N_{BB}}{N_{BA}+N_{BB}} \nonumber
    \end{eqnarray}
\item \underline{contamination of a class}: $cntN$. It is the dual of the purity, namely it is the ratio between misclassified object in a class and the number of objects classified in that class. In our confusion matrix example will be:
    \begin{eqnarray}
    cntA=1-pcA=\frac{N_{BA}}{N_{AA}+N_{BA}} \nonumber
    \end{eqnarray}
    \begin{eqnarray}
    cntB=1-pcB=\frac{N_{AB}}{N_{AB}+N_{BB}} \nonumber
    \end{eqnarray}
\end{itemize}

All these statistical indicators are packed in an output file, produced at the end of the test phase of any classification experiment.

%%%%%%%%%%%%%%%%%%%%%%%%%%%%%%%%%%%%%%%%
% SECTION 6.2
\section{Outliers analysis}
\label{out}
%%%%%%%%%%%%%%%%%%%%%%%%%%%%%%%%%%%%%%%%

For what the analysis of the catastrophic outliers is concerned, according to \cite{mobasher2007}, the parameter
$D_{95} \equiv \Delta_{95}/\left(1+z_{phot}\right)$ enables the identification of outliers in photometric redshifts derived through SED fitting methods (usually evaluated through numerical simulations based on mock catalogues). In fact, in the hypothesis that the redshift error $\Delta z_{norm} = \left( z_{spec}-z_{phot}\right)/\left(1 + z_{spec}\right)$ is Gaussian, the catastrophic redshift error limit would be constrained by the width of the redshift probability distribution, corresponding to the $95\%$ confidence interval, i.e. with $\Delta_{95} = 2\sigma \left( \Delta z_{norm} \right)$. In our case, however, photo-z are empirical, i.e. not based on any specific fitting model and it is preferable to use the standard deviation value $\sigma \left( \Delta z_{norm} \right) $  derived from the photometric cross matched samples, although it could overestimate the theoretical Gaussian $\sigma$, due to the residual spectroscopic uncertainty as well as to the method training error. Hence, according to the most popular rule, we consider as catastrophic outliers the objects with $\left| \Delta z_{norm} \right| > 0.15$.

It is also important to notice that for empirical methods it is useful to analyze the correlation between
the $NMAD\left( \Delta z_{norm} \right) = 1.48 \times median \left( \left| \Delta z_{norm} \right| \right)$ and the standard deviation $\sigma_{clean}(\Delta z_{norm})$ calculated on the data sample for which $\left| \Delta z_{norm} \right| \leq 2 \sigma \left( \Delta z_{norm} \right)$. In fact, it is normally expected that the quantity $NMAD$ would be less than the value of the $\sigma_{clean}$. In such condition we can assert that the pseudo-gaussian distribution of $\left( \Delta z_{norm} \right)$ is mostly influenced by the presence of outliers.

\begin{figure}[t]
\centering
\includegraphics[width=\textwidth]{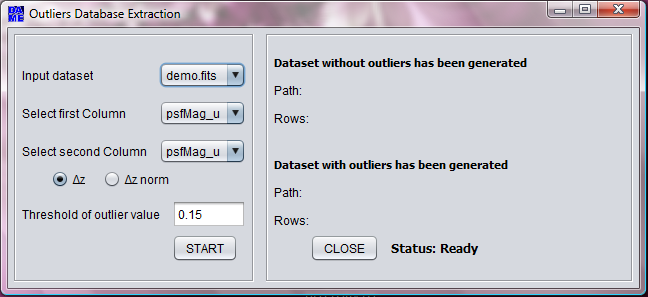}
\caption[Outliers window]{\footnotesize Outliers analysis setup window.}
\label{fig:outliers}
\end{figure}

Like for the statistical indicators, PhotoRApToR has also a tool to check the presence of outliers in local datasets.
For this reason it is useful to rewrite the relations for $\Delta z$ and $\Delta z_{norm}$ in a more general way:
\begin{eqnarray}
 \Delta z &=& Col_1 - Col_2 \nonumber\\
 \Delta z_{norm} &=& \frac{Col_1 - Col_2}{1 + Col_1} \nonumber
\end{eqnarray}

By clicking on \textbf{Regression}$>$\textbf{Outliers} a window is opened, where the user can select the dataset to check, and two drop down menus appear (Fig.~\ref{fig:outliers}), to select the \textit{first} and  the \textit{second} columns between which to estimate $\Delta z$ or $\Delta z_{norm}$. The text field below allows to set the threshold over which to consider an object as an outlier: if this tool is used after a regression experiment, the user can assign such threshold equal to the value of $\sigma$, as given by the statistics report, or the traditional value $0.15$.\\
The \textbf{START} button enables the outlier analysis. The tool generates a subset without outliers and one with only the outliers and finally, in the right panel, the path of the two files and the number of objects in each subset are reported.

%%%%%%%%%%%%%%%%%%%%%%%%%%%%%%%%%%%%%%%%
% SECTION 6.3
\section{Plotting tools}
\label{plot}
%%%%%%%%%%%%%%%%%%%%%%%%%%%%%%%%%%%%%%%%

Besides statistics, the PhotoRApToR application makes available also some graphical tools, useful to perform a visual inspection of any experiment. In particular a $2D$ scatter plot to show the trend of photo-z vs zspec, as well as histograms to graphically evaluate the distributions of quantities $\Delta z$ and $\Delta z_{norm}$.\\

Within the PhotoRApToR application there are present also instruments to generate different types of plot. These options are particularly suited during the preparation phase of data for experiments. They in fact enable the possibility to inspect trends, to quantify specific subsets of data, to compare distributions, as well as to inspect particular trends of a generic data table. The graphical options selectable by user are:
\begin{itemize}
\item multi-column histograms;
\item multiple 2D plots;
\item multiple 3D scatter plots.
\end{itemize}

When one of the plot options (Histo Plot, Scatter Plot or 3D Plot) is clicked, a new window is opened (Figures \ref{fig:histo}, \ref{fig:scatter} and \ref{fig:3d}) where to set the plot parameters: in the upper panel will be displayed the plot, below there are a text field where it is possible to set the name of the plot and two
checkboxes that allow to enable/disable a grid and a legend for the plot. The \textbf{Add Plot} button adds other tabs to the previous panel where it is possible to set the parameters of the combined plot with different colours in such a way to compare data from different tables.\\
By clicking on the \textbf{Plot} button, in the upper panel the plot is displayed and stored on local directory in JPEG file format.

%%%%%%%%%%%%%%%%%%%%%%%%%%%%%%%%%%%%%%%%
% SECTION 6.3.1
\subsection{Histo Plot}
\label{histo}
%%%%%%%%%%%%%%%%%%%%%%%%%%%%%%%%%%%%%%%%

\begin{figure}[h]
\centering
\includegraphics[width=\textwidth]{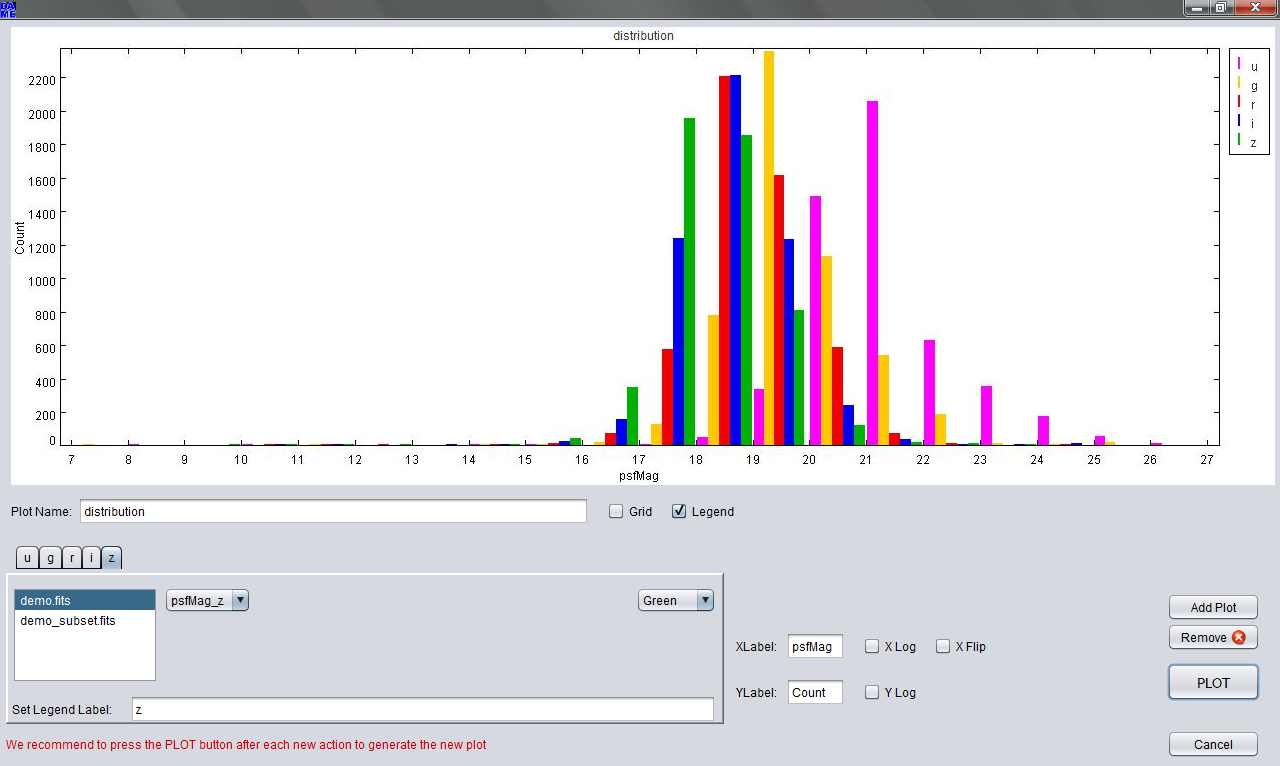}
\caption{\footnotesize Histogram (Histo Plot option) panel.}
\label{fig:histo}
\end{figure}

Each plot option has a different panel where to set parameters. For the \textbf{Histo Plot} (Fig.\ref{fig:histo}):
\begin{itemize}
 \item a \textbf{Table List} that is the same of the main window;
 \item a drop-down menu to set the \textbf{X-axis} of the diagram;
 \item two text fields where it is possible to change the labels for the axes X and Y;
 \item two checkboxes for each axis, one to flip and another to set the axis in logarithmic scale;
 \item another drop-down menu allows to set the colour;
 \item there is also another text field where to change the label for the plot legend.
\end{itemize}

%%%%%%%%%%%%%%%%%%%%%%%%%%%%%%%%%%%%%%%%
% SECTION 6.3.2
\subsection{Scatter Plot}
\label{scatter}
%%%%%%%%%%%%%%%%%%%%%%%%%%%%%%%%%%%%%%%%

\begin{figure}[h]
\centering
\includegraphics[width=\textwidth]{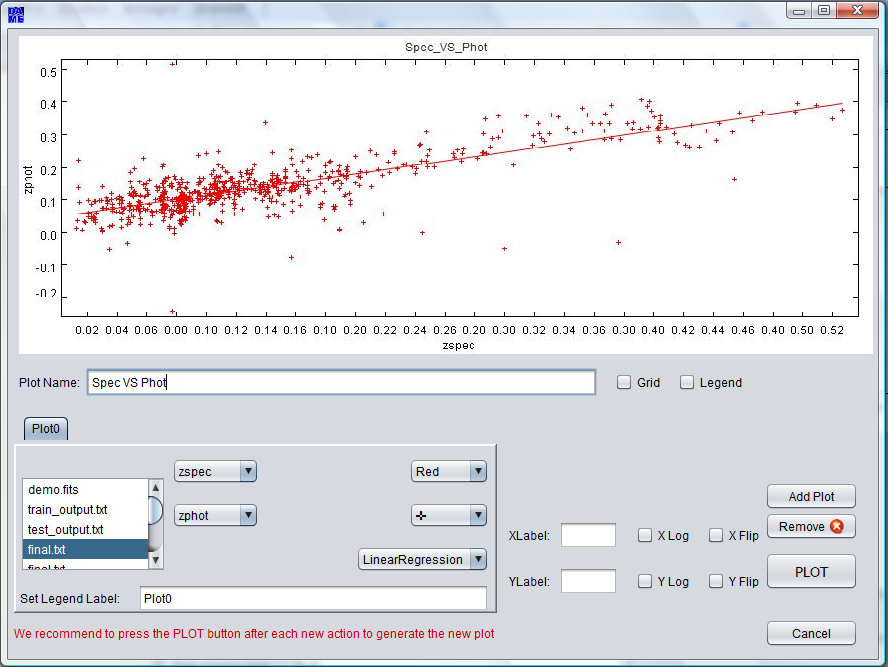}
\caption{\footnotesize Scatter Plot panel.}
\label{fig:scatter}
\end{figure}

For the \textbf{Scatter Plot} option (Fig.\ref{fig:scatter}) there are:
\begin{itemize}
 \item two drop-down menus to set the \textbf{X-axis} and \textbf{Y-axis} of the diagram;
 \item two text fields where it is possible to change the labels for the axes X and Y;
 \item two checkboxes for each axis, one to flip and another to set the axis in logarithmic scale;
 \item three drop-down menus allowing to set the \textit{Line Style}, the \textit{Colour} and the \textit{Marker};
 \item there is also another text field where to change the label for the plot legend.
\end{itemize}

%%%%%%%%%%%%%%%%%%%%%%%%%%%%%%%%%%%%%%%%
% SECTION 6.3.3
\subsection{3D Plot}
\label{3d}
%%%%%%%%%%%%%%%%%%%%%%%%%%%%%%%%%%%%%%%%

\begin{figure}[h]
\centering
\includegraphics[width=\textwidth]{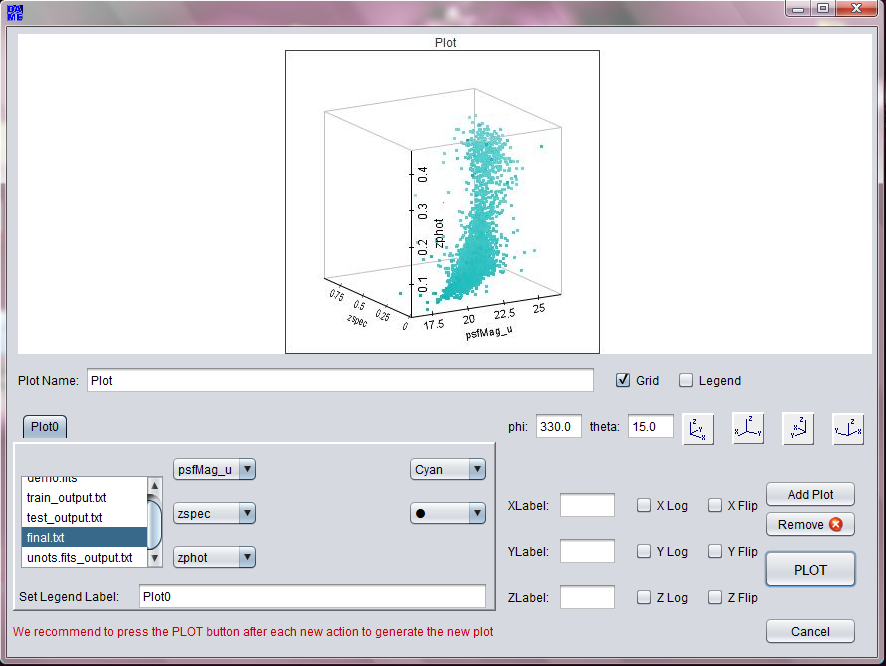}
\caption{\footnotesize 3D Plot panel.}
\label{fig:3d}
\end{figure}
Finally, for \textbf{3D Plot} (Fig.\ref{fig:3d}):
\begin{itemize}
 \item three drop-down menus to set the \textbf{X-axis}, \textbf{Y-axis} and \textbf{Z-axis} of the diagram;
 \item three text fields where it is possible to change the labels for the axes X, Y and Z;
 \item two checkboxes for each axis, one to flip and another to set the axis in logarithmic scale;
 \item two drop-down menus allowing to set the \textit{Colour} and the \textit{Marker};
 \item there is also another text field where to change the label for the plot legend.
\end{itemize}

%%%%%%%%%%%%%%%%%%%%%%%%%%%%%%%%%%%%%%%%
% CHAPTER 7
\chapter{Troubleshooting}
\label{trouble}
%%%%%%%%%%%%%%%%%%%%%%%%%%%%%%%%%%%%%%%%

This chapter provides answers to specific troubles arising from the use of the application. In order to improve your user experience with PhotoRApToR, it is recommended to read this section to learn more about common pitfalls and to
get recommendations on how to correctly use the application and to recover wrong situations as well.\\

For any request the user can send an e-mail to \textit{helpdame@gmail.com} and will be re-contacted as soon as possible.

%%%%%%%%%%%%%%%%%%%%%%%%%%%%%%%%%%%%%%%%
% SECTION 7.1
\section{Heap memory limit}
\label{mem}
%%%%%%%%%%%%%%%%%%%%%%%%%%%%%%%%%%%%%%%%

During the Pre-processing phase, the user must check that the dimension of the loaded dataset file is smaller than the JVM assigned memory.\\

The Java heap is where the objects of a Java program lives. It is a repository for alive and dead objects, and free memory as well. When an object can no longer be reached from any pointer in the running program, it is considered \emph{garbage}.
If the dataset is not allocated in the Java heap, this error does not necessarily imply a memory leak. The problem can be as simple as a configuration issue, where the default heap size is insufficient for the application.\\

In these cases, users must specify a new Java heap size values. This can be done executing PhotoRaptor from a terminal with a command line as follow:\\

\textbf{java -jar -Xms***m -Xmx***m PhotoRApToR.jar} \\

that allows to allocate *** megabytes of memory to the minimum (-Xms) and maximum (-Xmx) heap sizes. The default size for these values is measured in bytes. Append the letter `k' or `K' to the value to indicate kilobytes, `m' or `M' to indicate megabytes, and `g' or `G' to indicate gigabytes.

%%%%%%%%%%%%%%%%%%%%%%%%%%%%%%%%%%%%%%%%
% SECTION 7.2
\section{Corrupted datasets}
\label{corr}
%%%%%%%%%%%%%%%%%%%%%%%%%%%%%%%%%%%%%%%%

PhotoRApToR allows to open and to edit datasets in different file formats, but when users try to load a file, the application checks if data are not corrupted, or loaded as a wrong type specification.\\
A warning dialog is shown if the file is wrong or corrupted and it is not added to the Table List.\\

One of most frequent cases of a corrupted dataset happens when users edit data tables ignoring table metadata.

%%%%%%%%%%%%%%%%%%%%%%%%%%%%%%%%%%%%%%%%
% SECTION 7.3
\section{Filenames with spaces}
\label{space}
%%%%%%%%%%%%%%%%%%%%%%%%%%%%%%%%%%%%%%%%

Datasets whose name contains spaces or that are located in folders whose name contains spaces, generate errors if used during the Experiment phase.\\
If users try to use this type of file as input for an experiment or if the name chosen for the output folder contains spaces, PhotoRApToR will show a warning dialog and the operation is stopped.\\

It is necessary to use data folders whose names are specified without spaces and to control dataset names before to run PhotoRApToR.

%%%%%%%%%%%%%%%%%%%%%%%%%%%%%%%%%%%%%%%
% SECTION 7.4
\section{Mac OS X window focus problem}
\label{macfocus}
%%%%%%%%%%%%%%%%%%%%%%%%%%%%%%%%%%%%%%%%

There could be occasional incompatibilities between the Java Swing tool and Mac OS X systems. Within the application this may primarily occur with a particular panel: after having selected the photo-z Menu button, the new setup window may result not editable. In such situation it is suggested to change focus, by clicking on any other open window in your desktop and then to select again the photo-z setup panel. Hereinafter the user should be able to proceed with the normal setup.

%%%%%%%%%%%%%%%%%%%%%%%%%%%%%%%%%%%%%%%%
% SECTION 7.5
\section{Generic problems}
\label{generic}
%%%%%%%%%%%%%%%%%%%%%%%%%%%%%%%%%%%%%%%%

\begin{itemize}
\item \textbf{\textit{After installation the program doesn't start}}: the most common reason could be the absence or wrong setup of the JVM on the user machine. Please check it and in case download and install the latest version of the JVM compatible with the OS running on the user machine. The official website is \url{http://www.oracle.com/technetwork/java/index.html}. Another source of failure could be the wrong downloaded version of the package, not matching the user OS running on the local machine. Please verify carefully this requirement. In case of still wrong execution of the program, please contact us by specifying the wrong condition/message.
\end{itemize}

%--------------------------------------
\appendix
\chapter{The Machine Learning model}
\label{mlpqna}
%--------------------------------------

As introduced, the core engine of the PhotoRApToR application is the ML model underlying all data mining experiments, for instance the MLPQNA method. It is a Multi Layer Perceptron (MLP; \citealt{rosenblatt1961}) neural network trained by a learning rule based on the Quasi Newton Algorithm (QNA). It is one of the widely used feed-forward neural networks in a large variety of scientific and social contexts.

The Quasi Newton Algorithm (QNA) is a variable metric method for finding local maxima and minima of functions \citep{davidon1991}. The model based on this learning rule and on the MLP network topology is then called MLPQNA. QNA is based on Newton's method to find the stationary (i.e. the zero gradient) point of a function. The QNA is an optimization of Newton based learning rule, because the implementation is based on an incremental approximation of the Hessian by a cyclic gradient calculation. In PhotoRApToR the Quasi Newton method has been implemented by following the known L-BFGS algorithm (Limited memory - Broyden Fletcher Goldfarb Shanno; \citealt{byrd1994}). As a matter of fact, this method was designed to optimize the functions with a variable number of arguments (hundreds to thousands), because in this case it is worth to have an increased number of iterations, due to the lower approximation precision. This is particularly useful in astrophysical data mining problems, where usually the parameter space is dimensionally huge and is often affected by a low signal-to-noise ratio.

Most of the analytical characteristics of the method have been deeply described in the contexts of both classification \citep{brescia2012b} and regression \citep{brescia2013a,cavuoti2012b}. We suggest to refer to these articles in case of interest about mathematical theory behind the method as well as on examples of its applications in real astrophysical contexts.

\section*{Acknowledgments}

\noindent The authors wish to thank the whole DAME working group for their fundamental contribution.
\noindent MB wishes to thank the financial support of PRIN-INAF 2010, \textit{Architecture and Tomography of Galaxy Clusters} and the PRIN-INAF 2014 {\it Glittering kaleidoscopes in the sky: the multifaceted nature and role of Galaxy Clusters}.
\noindent MB and SC acknowledge financial contribution from the agreement ASI/INAF I/023/12/1 for the Euclid space mission.
\noindent The authors also wish to thank the financial support of Project F.A.R.O. III Tornata (University Federico II of Naples).


\bibliographystyle{model2-names}
%% not want to use model2-names.bst.


\end{document}